\documentclass [12pt]{article}
\usepackage {amssymb}
\usepackage {amsmath}
\usepackage {graphicx}
\usepackage {longtable}

\begin{document}

\begin{flushleft}\footnotesize{ {\it Kybernetes: The International Journal
of Systems \& Cybernetics} {\bf 32}, No. 7/8, pp. 1005-20 (2003)
{\rm (a special issue on new theories about time and space, Eds.:
L. Feng, B. P. Gibson and Yi Lin)}}
\end{flushleft}

\section*{Scanning the structure of ill-known spaces: Part 3.  \\
Distribution of topological structures at elementary and cosmic scales}

\medskip

\begin{center}
\textbf{Michel Bounias$^{(1)}$ and Volodymyr
Krasnoholovets$^{(2)}$}
\end{center}

\begin{center} ($^{1}$) BioMathematics Unit(University/INRA, France, and
IHS, New York, USA), \\ Domain of Sagne-Soulier, 07470 Le Lac
d'Issarles, France            \\
($^{2}$) Institute of Physics, National Academy of Sciences, \\
Prospect Nauky 46,     UA-03028, Ky\"{\i}v, Ukraine
\end{center}

\textbf{Abstract.} The distribution of the deformations of
elementary cells is studied in an abstract lattice constructed
from the existence of the empty set. One combination rule
determining oriented sequences with continuity of set-distance
function in such spaces provides a particular kind of
spacetime-like structure that favors the aggregation of such
deformations into fractal forms standing for massive objects. A
correlative dilatation of space appears outside the aggregates. At
the large scale, this dilatation results in an apparent expansion,
while at the submicroscopic scale the families of fractal
deformations give raise to families of particle-like structure.
The theory predicts the existence of classes of spin, charges and
magnetic properties, while quantum properties associated to mass
have previously been shown to determine the inert mass and the
gravitational effects. When applied to our observable spacetime,
the model would provide the justifications for the existence of
the creation of mass in a specified kind of "void", and the
fractal properties of the embedding lattice extend the phenomenon
to formal justifications of Big-Bang-like events without need for
any supply of an extemporaneous energy.

\bigskip
\textbf{Key words:} continuity, distributions, fractal quanta,
mass creation, particle families

\bigskip
\textbf{PACS classification:} 02.10.Cz -- set theory, 02.40.Pc --
general topology, 03.65.Bz -- foundation theory of measurement,
miscellaneous theories

\newpage

\section*{1. Introduction}

\hspace*{\parindent} Despite the striking progress in present-day
research, from corpuscle physics (Fritzsh, 2000) to astrophysics
(B\"orner, 2000), many fundamental questions remain unsolved and
often contradictory (Krasnoholovets, 2001).

In previous papers (Bounias and Krasnoholovets, 2003a,b) formal
demonstrations have identified mass with a disruption in
homeomorphic mappings of reference medium, from one to the next
Poincar${\rm \acute{e}}$ section whose ordered sequence stands for
a spacetime-like structure (Bonaly and Bounias, 1995). In an
attempt to identify forlam conditions of existence of a
physical-like world, the existence of the empty set as the
founding space alongwith theory of sets and topology, extended to
nonwellfounded sets as the combination rules, was found as
necessary and sufficient conditions (Bounias and Bonaly, 1997;
Bounias, 2000).

This paper starts from the lattice of empty elements which are
balls constituted from the empty set and its successive
complementaries, all exhibiting self-similarity and fractal
properties (Bounias and Krasnoholovets, 2003a,b). Elementary balls
within a given range of size were attributed a virtual volume at
the free state. These volumes in fact are reference frames in
which the position of objects is assessed by an operator called
the "moment of junction", since it connects one to the next
Poincar$\acute{\rm e}$ sections and owns the structure of a moment
(Bounias, 1997). These volumes belong to the space of distances
(topologically open), as a topological complementary of the space
of objects (topological closed), and they remain belonging to this
class as far as their morphisms are homeomorphic. In contrasts,
balls exhibiting dimensional changes (here through fractal
shaping) no longer fulfil this condition, and they have been
attributed to the class of objects (Bounias and Krasnoholovets,
2003b). However, at this stage, no rationale was yet provided for
the justification of existence of such structures: this point will
therefore be addressed in first in this paper.

\section*{2. Preliminaries}

\hspace*{\parindent} The distribution of variables X, Y is a
density function h(x,y) which admits for margin densities for each
variable the following (R\"{u}egg, 1985):
 $$
\rm f(x) = \int h(x,y) d x \ \ and \ \ g(y) = \int h(x,y) d y.
 \eqno(1.1)
 $$

If E is a probabilized space, which is the case of the topological
spaces in which we are working (Bonaly and Bounias, 1995, Bounias,
2000), and the variables are continued, then the probability that
x, y belong to E is:
 $$
\rm P(x,y \in E) = P(E) = {\iint\limits_{E}} h(x,y) d x d y.
 \eqno(1.2)
 $$

For discrete variables, the integral is replaced by a union or a sum. The
repartition function is represented by a summation (again in either sense)
with boundaries:
 $$
\rm H(x,y) = P\{(X \leq x)\cap (Y \leq y)\}
 \eqno(1.3)
 $$

More generally, one may consider $\rm x \in [u,{\kern 2pt} v]$
within a domain $\rm [a,{\kern 2pt} b]$ of E. Then the probability
of finding x in the closed segment $\rm [u,{\kern 2pt}v]$ is $\rm
P(u < x < v) = (v-u)/(b-a)$ if E is totally ordered. In other
cases, alternative solutions have been examined in Bounias and
Krasnoholovets, (2001a). Then the process will be extended to $\rm
y \in [q,{\kern 2pt}r]$, and so on. In a discrete space,
probabilities are multiplicative, while in a continued space the
repartition functions are multiplicative: $\rm H(x,y,...) =
F(x)\cdot G(y) \cdot (...)$.

Now, let X, Y,... be random objects defined on the same
probabilized space, and $\rm Z = O(X,{\kern 1pt}Y,{\kern 1pt}...)$
a real function in E. Then the moment, including the expected
value, wears the form: $\rm E(Z) = \iint ... O(X,Y,...)\cdot
h(x,y,...)\cdot dx{\kern 1pt}dy(...)$.

Whatever the form of a distribution, it owns a family of moments
$\mathfrak{m}^{0}_1$ of order k and centered on c: the expected
value is $\rm E = \mathfrak{m}^0_1$, and the variance $\rm Var =
\mathfrak{m}_2^E$.

One particular case is the covariance $\rm Cov(X,{\kern
1pt}Y)=E(X,Y) - E(X)\cdot E(Y)$, so that if one has $\rm
O(X,{\kern 1pt}Y,{\kern 1pt}...) = X+Y$, then:
 $$
\rm Var(X+Y) = Var(X) + Var(Y) + 2{\kern 1pt} Cov(X,Y)
 \eqno(1.4)
 $$

This brings the question of the dependence of variables X and Y:
$\rm Cov(X,Y)$ is bounded by zero for X,Y independent and by a
maximum if X and Y are completely self-similar, like in any
subpart of a fractal structure.

Finally, the distribution K(z) as the probability to get the sum
$\rm (X+Y +...) \leq z$ is given by the derivative of the
repartition:
 $$
\rm P(X+Y+...\leq z) = \iint\limits_{x+y+ ... \leq z} ...
h(x,y,...)\cdot dx{\kern 1pt} dy{\kern 1pt}(...).
 \eqno(2.1{\kern 1pt} \rm a)
 $$
The summation on one variable, e.g. y, is bounded by $\rm
z-(x+...)$
 $$
\rm P(X + Y + ... \leq z) = K(z) =
\int\limits_{-\infty}^{+\infty}...\big(f(x)...\big)
\Big(\int\limits_{-\infty}^{z-(x+...)}g(y)\cdot dy \Big) \cdot dx
\eqno(2.1 \rm b)
 $$
that is in terms of distribution:
 $$
\rm k(z) =
\int\limits_{-\infty}^{+\infty}...\big(f(x)...\big)\big(g(z-(x+...))\big)
dx
 \eqno(2.2{\kern 1pt}\rm a)
 $$
that is a convolution function.

\medskip \noindent
\textbf{Remark 2.1.} We have shown in Part 2 of this study that
the morphisms of distances and objects already fulfil a nonlinear
form of generalized convolution:
 $$
\rm (\mathcal{M}\perp\mathcal{J})_{{\kern 1pt}
k+i}=T^{\perp}(\mathcal{M} \bigcirc \mathcal{J})_{{\kern 1pt}k}
\eqno(2.2{\kern 1pt}\rm b)
 $$
where $\mathcal{M}$ and $\mathcal{J}$ are morphisms of distances
and objects, respectively, and T an operator mapping a
Poincar\'{e} section (Si) into (Si{\small{+{\kern 1pt}1}}), on the
basis of the moment of junction MJ, that is a composition function
of either the set distances or their complementaries (the
"instans") with a distribution function (Bounias, 1997). The
operator T translates a composition rule (here: $\bigcirc$) into
($\perp$).

Redundancy will be considered in either active or with commutation
forms. The latter involves multiple convolution of densities:
 $$
\rm f^{{\kern 1pt} n^*} =f{\mbox{\scriptsize 1}}\ast
f{\mbox{\scriptsize 2}} \ast ...\ast f{\mbox{\scriptsize i}}\ast
...\ast f{\mbox{\scriptsize n}}.
 \eqno(2.3)
 $$

\section*{3. Main results}

\subsection*{3.1. On the law determining the sequence of Poincar\'{e}
sections}

\hspace*{\parindent} Let Si be a closed intersection of
topological dimension n produced by the intersection of a
n-subspace with a m subspace ($\rm m>n$) belonging to the set of
parts of the embedding $\omega$-space (W$^{{\kern 1pt}\omega}$).

A universe will thus be constructed in a space ($\rm W^{{\kern
1pt}\omega}) = \{X, \perp \}$ with $\rm X \in \{X^3 \cap X^4 \}$ a
set and ($\perp$) a combination rule determining the choice of
Si{\small{+{\kern 1pt}1}} from Si.

\medskip \noindent
\textbf{Remark 3.1.} It has been argued (Bonaly and Bounias, 1995)
that $ \omega = 4$ and $\rm n=3$ provide an optimal situation, in
terms of mathematical organizational properties, which would place
our spacetime among the most efficient universe configurations.
Thus, throughout this study, it will be sufficient to consider
$\rm n=3$ coming from $\omega =4$.

\medskip \noindent
\textbf{Remark 3.2.} There exists as many universes as there are
laws ($\perp$). However, one particular case deserves particular
attention.

\medskip \noindent
\textbf{Proposition 3.1.} Let Si denote a 3-D Poincar\'{e} section
of W$^{4}$. Continuity in mappings of members of Si in the
sequence \{Si\}$_{{\kern 0.3pt}\rm i}$ is favoured if the
successors Si{\small{\kern 1pt+1}} are such that:
 $$
\rm \{Si\}_{i} = (\forall {\kern 1pt} Si, {\kern 2pt} Si \cap Si+
\! \mbox{\footnotesize{ 1}} = max\{Si \cap Si+k \}\! _{{\kern 2pt}
\rm k>i}).
 $$

\medskip\noindent
\underline{Proof.} Some lemmas of continuity of set-distance
functions will first be demonstrated, and the proposition will
then be deduced.

\subsubsection*{3.1.1. Continuity of set distance functions}

\hspace*{\parindent} The following definition recalls the
generalized distance provided by topologies as it has been
presented in Part 1 (Bounias and Krasnoholovets, 200la).

\medskip \noindent
\textbf{Definition 3.1.} Let E be a topological space $\rm  E=
\{X,{\kern 1pt} T\}$, and A, B, C, ..., G, ... subspaces
constituted from the set of parts of set X composing E. Then, the
separating set-distance between A and B within E is denoted by
$\rm \Lambda_E(A,B)$ and identified by:
 $$
\rm \Lambda_E(A,B) = min \Big\{(G\in E), {\kern 2pt}(A\in G \neq
0, {\kern 2pt} B\in G\neq 0): {\kern 1pt}\Delta (A,G) \cap \Delta
(B,G) \Big\}
 \eqno(3.1)
 $$
where $\Lambda$ denotes the simple set-distance as the symmetric
difference:
 $$
\rm \Delta(A,B) = {\mathop{\complement}\limits_{A {\kern 1pt} \cup
{\kern 1pt} B}} (A \cap B).
 \eqno(3.2)
 $$

The generalized set-distance if given by the following relation:
 $$
\rm \Delta_{E}(A,B) = min \Big \{ (G\in E), {\kern 1pt}(A\cap
G\neq {\O}, {\kern 2pt} B\cap G \neq {\O}): {\kern 2pt}\Delta
(A,B,G) \}
 \eqno(3.3)
 $$
with
 $$
\rm \Delta(A,B,G) = {\mathop{\complement} \limits_{A\cup {\kern
1pt} B {\kern 1pt}\cup {\kern 1pt} G}} \Big( (A \cap B) \cap (A
\cap G)\cap (B\cap G) \Big).
 \eqno(3.4)
 $$

If $\rm G \neq ${\O}, then relation (3.3) reduces to (3.2) and
$\rm \Delta_E$ reduces to $\Delta$.

\medskip \noindent
\textbf{Lemma 3.1.} The mapping $\rm {\it f}\!\!: {\kern 1pt}
\Delta \mapsto R$ of the set distance ($\Delta$) on the set of
real numbers ($\mathbf{R}$) is continuous.

\medskip \noindent
{\underline{Proof.} Let A, B in E be mapped into $\rm {\it
f}(A)=a$ and $\rm {\it f}(B)=b$ in {\bf R}. If a and b are
cuttings, the proof is trivial. If a and b are initial segments
(like simple numbers) then, take the case where $\rm a < b$, and
consider e as small as needed, such that $\rm a^\prime = a+e$. For
any e, there exists x in E such that $\rm e = {\it f}(x)$. When
the distance $\rm \delta \{\Delta(A,B), {\kern 2pt}(A^\prime,B)\}$
is decreased by x, then the difference $\rm (b-a)$ becomes $\rm
((b-a)-e)$, i.e. it is decreased by e.
\qquad\qquad\qquad\qquad\qquad\qquad \ \ \ (Q.E.D.)

\medskip \noindent
\textbf{Lemma 3.2.} Let A, B, G in E and $\rm {\it f}(A)=a$, $\rm
{\it f}(B)=b$ and $\rm {\it f}(G)=g$ in {\bf R}. \newline (i) The
mapping $\rm { \! \mathfrak{L}\!\!\! \mbox{-}}{\kern 1pt}:{\kern
2pt}\Lambda_E \mapsto R$ of the separating distance on the set of
real numbers is continuous if a, b, g are cuttings. \newline (ii)
If a, b, g are initial segments, the mapping remains continuous if
E is totally ordered, while if E is only partly ordered by
inclusion or intersection, then the mapping ${ \!
\mathfrak{L}\!\!\! \mbox{-}}${\kern 1pt} is continuous for any
$\rm e<a$ or $\rm e>b$.

\medskip  \noindent
\underline{Proof.} The first case is trivially infering from the
continuity of A. In the second case, if $\rm a<e<b$, then $\rm
dist(a,{\kern 0.4pt}g)$ and $\rm dist(b,{\kern 0.4pt}g)$ have a
null difference only if $\rm g= \langle a,b \rangle$ or if they
were to be considered as adjacent cuttings: then their
intersection would always be null. However, in these two
particular cases, the mapping £ remains correct if E is totally
ordered, so that $\rm A \subset G \subset B$ and $\rm \Lambda_E =$
{\O}. Then continuity is proved for any (g).

\subsubsection*{3.1.2. Continuity in ordered Poincar\'{e} sections of
space}

\hspace*{\parindent} Let ($\rm S_{{\kern 1pt}i}$) be one 3-D
timeless section in W$^{4}$, {\kern 1pt} $\rm a_{{\kern 0.5pt} i}$
be a member or a part of ($\rm S_{{\kern 0.5pt}i}$) and $\rm
\mathcal{V}(a_{{\kern 0.5pt}i})$ a neighborhood of $\rm (a_{{\kern
0.5pt}i})$ in ($\rm S_{{\kern 0.5pt}i}$). Call $\rm (\underline{
a}_{{\kern 0.5pt}i})_{{\kern 1pt}i+k}$ and $\rm
\underline{\mathcal{V}}(\underline{a}_{{\kern 0.5pt}i})_{i+k}$ the
homeomorphic projections of $\rm a_{{\kern 0.5pt}i}$ and $\rm
\mathcal{V}(a_{{\kern 0.5pt}i})$ on (S$_{{\kern 0.5pt}\rm i+k}$).
Proposition 1 states that $\rm \Delta\big((S_{{\kern
0.5pt}i},(S_{{\kern 0.5pt}i+1})\big)$ must be minimal and that for
the same reason, $\rm \Delta \Big( \big(
\underline{\mathcal{V}}(\underline{a}_{{\kern 1pt}i}) \big)_{i+1},
{\kern 1pt}\big(\mathcal{V}(a_{{\kern 1pt}i+1})\big)\Big)$ is
minimal, which is consistent with the clause of continuity. If, in
contrast, there exists a section ($\rm S_{{\kern 1pt}i+h}$) whose
distance with ($\rm S_{{\kern 1pt}i}$) is smaller than $\rm
\Delta\big((S_{{\kern 1pt}i}),{\kern 1pt}(S_{{\kern
1pt}i+1})\big)$, then the neighborhood $\rm
\big(\mathcal{V}(a_{{\kern 1pt}i+1})\big)$ may be contained in
$\rm \Delta \big( (S_{{\kern 1pt}i}) \cap (S_{{\kern 1pt}i+h}),
{\kern 1pt} (S_{{\kern 1pt}i+1})\big)$.

In particular, one may have
 $$
\rm \Delta_{E}\Big(\big(\underline{\mathcal{V}}(a_{{\kern
1pt}i})\big)_{{\kern 0.5pt}i+1}, {\kern 1pt}
\big(\mathcal{V}(a_{{\kern 1pt}i+1})\big) \Big)\Delta\supset \rm
\Big( \big( {\underline{\mathcal{V}}}({\underline{a}}_{{\kern
1pt}i})\big)_{{\kern 1pt}i+h},{\kern 2pt}\big(
\mathcal{V}(a_{{\kern 1pt} i+h}) \big) \Big)
 $$
and the condition of continuity is no longer necessarily
fulfilled. This achieves the justification.

\subsection*{3.2. Distribution of the deformations of lattice balls}

\subsubsection*{3.2.1. Introduction}

\hspace*{\parindent}  Sections $\rm \{S_{{\kern 1pt}i}\}_{i}$ are
composed as pointed in Part 1 of distance ($\Delta$ and $\Lambda$)
and objects ($\mathfrak{m}\langle {\kern 1pt}{\kern 1pt}\rangle$).
The former are open and the second are closed. The space of
distances provides the reference frame from which the topological
changes of objects localization will be observed. This space has
been shown in Part 2 to be basically constituted of elementary
cells represented by free forms C$^{{\kern 1pt}\rm free}$ and
degenerate forms C$^{{\kern 1pt}\rm deg}$. A putative volume $\rm
V^{free} = V_\circ$ is attributed to free cells which are devoid
of any deformations and thus described by the identity mapping
(Id) from (Si) to ($\rm Si+1$). In contrast, degenerate cells
result from homeomorphic transformations, which involve some
change in their volumes (in 3-D sections) without dimensional
alteration. Then, if $\rm \delta V_\circ$ canonically denotes such
a change in the volumes, then $\rm V^{deg} = V_\circ \pm \delta
V_\circ)$. From sections (Si) to $\rm (Si+1)$ one has $\rm \delta
V_{\circ{\kern 1pt} (i)}$ mapped into $\rm \delta V_{\circ{\kern
0.5pt}(i+1)}$. Within each section, the set of all such
deformations will be $\rm \cup_{{\kern 1pt}i}{\kern 1pt}
\{C_{{\kern 1pt}i}^{{\kern 1pt} deg}\}_i$ and $\rm \cup_{{\kern
1pt}i+1}{\kern 1pt}\{ C_{{\kern 1pt}i}^{{\kern 2pt} deg}\}_{i+1}$
respectively.

\medskip \noindent
\textbf{Remark 3.3.} The distribution of these cells within each
section will concern as many variables and will be decribed by a
multiple convolution as in relation (2.3).

\subsubsection*{3.2.2. Distribution of deformations}

\hspace*{\parindent} Now, consider the fate of the homeomorphic
projections (\underline{C}$\rm ^{{\kern 1pt} deg}_{{\kern
2pt}i})_{{\kern 1pt}i+1}$ and (C$\rm ^{{\kern 1pt}deg}_{{\kern
1pt}i+1})$. According to relation (1.4), one has respectively:
  $$
\rm Var \big((\underline{\delta}^{{\kern 1pt} deg}_{{\kern 2pt}
i})_{{\kern 1pt} i+1} + (\delta_{{\kern 1pt} i+1}^{{\kern 1pt} deg
})\big) = Var{\kern 1pt}\big((\underline{\delta}^{{\kern 1pt}
deg}_{{\kern 2pt} i})_{{\kern 1pt} i+1} \big) + Var{\kern
1pt}(\delta_{{\kern 1pt} i+1}^{{\kern 1pt} deg }) + 2
Cov\big((\underline{\delta}^{{\kern 1pt} deg}_{{\kern 1pt}
i})_{{\kern 1pt} i+1}, {\kern 2pt} (\delta_{{\kern 1pt}
i+1}^{{\kern 1pt} deg })\big) \eqno(4.1)
 $$
and with $\rm N^{{\kern 1pt} deg}$ the cardinal (Card) of set $\rm
\{C_{{\kern 1pt}i}^{deg}\}$:
 $$
\rm Var\big( \cup _{{\kern 1pt} i}(\delta_{{\kern 1pt}i}^{{\kern
1pt} deg})\big) = \cup_{{\kern 1pt}i}{\kern 1pt} Var \big(
(\delta_{{\kern 1pt}i}^{{\kern 1pt}deg})\big) + N^{{\kern 1pt}deg}
\cdot Cov \big(\{ \delta_{{\kern 1pt}i}^{{\kern 1pt}deg}
\}_{{\kern 1pt} i}\big).
 \eqno(4.2)
 $$

\subsubsection*{3.2.3. Boundaries}

\hspace*{\parindent} Gather relations (4.1) and (4.2) in:
 $$
\rm Var{\kern 1pt}(\cup{\kern 1pt}(\delta)) = \cup {\kern 2pt}
Var(\delta) + Card(\cup) \cdot Cov(\{\delta,\}).
 \eqno(4.3)
 $$
These variances are subjected to boundary conditions, depending on
the level of dependence or independence of ($\rm \delta^{{\kern
1pt} deg}_{{\kern 2pt}i}$) and ($\rm \delta^{{\kern
1pt}deg}_{{\kern 1pt}i}$).   \newline  \noindent  \underline{First
kind.} {\kern 1pt} If the variables are totally independent, like
in a completely random space, one will get $\rm Cov(\{\delta,\})
=0$. Thus, the variance of the sum is minimal.  \newline \noindent
\underline{Second kind.} {\kern 1pt} In contrast, $\rm max{\kern
1pt}Cov(\{\delta,\})$ is attained if the components $\rm
(\delta{\kern 1pt}V_{\circ(i)}\}$ exhibit the maximum of
similarity. This condition is achieved through fractal properties
of the lattice, whose cells are self-similar balls composed with
the empty hyperset \{{\O}$^{\O}$\}.

\subsubsection*{3.2.4. Theorem of the distribution of volumes}

\hspace*{\parindent} Then, owing to Proposition 1, the selection
of (Si+{\small{1}}) from (S{\kern 0.2pt}i) will preferably retain
a first kind distribution.

\medskip \noindent
\textbf{Lemma 3.3.} The degenerate lattice contains a non
denumerable infinity of subdeformations.

\medskip
\noindent \underline{Proof.} It has been previously proved that
the empty hyperset provides existence of a n-space, n as great as
needed, endowed with the power of continuum (Bounias and Bonaly,
1997). Each empty set unit gives a empty complementary in itself,
so that each unit provides a sequence of structures fitted one
into the other, which can be indexed on a sequence of the $\rm
\{l/2^{{\kern 1pt} ni}\}_i$ type (Part 2). Thus, the distribution
of volumes $\rm \{ \delta V_{\circ (i)} \}$ in the degenerate
space contains infinitely many times the collections of
deformations required for constituting a quantum of fractality.
Hence, each time these quanta are available in the topological
neighborhood of a cell in (Si), the law of selection of the next
section will select (Si+{\small{1}}) in ($\rm W^{4}$) such that
the same set of deformation is organized into one single
structure, that is a fractal.

This now allows the following founding statement:

\medskip \noindent
\textbf{Corollary 3.1.} The combination rule of a continued
spacetime-like sequence of Poincar\'{e} sections fulfilling the
option stated in Proposition 1 exhibits a trends to collapsing
random distributions of degenerate cells into massive objects.

\medskip \noindent
\underline{Justification.} Continuity associated with the
condition of maximum intersection (Proposition 1) favour the
collapse of scattered deformations into one single aggregate
forming a fractal structure: this results in a change of
dimensionality of the affected cells. The latter are no longer
homeomorphic images, and therefore, they get a mass, in the sense
defined in Part 2. Therefore these cells escape the class of
"reference frame" or distances, and fall into the class of
"objects". They become "particled balls", denoted and their
volumes are $\rm V^{part}$ as described in Part 2.

\subsection*{3.3. Predicted structural classes of particled cells}

\subsubsection*{3.3.1. Predicted particle-like components}

\textbf{3.3.1.1. Mass-equivalent nonmassive corpuscles.} Denote by
$\rm \vartheta =\{\varrho,{\kern 1pt}a,{\kern 1pt}I\}$ a quantum
of fractality where
  $$\rm I = \sum_{{\kern 1pt}i=1 \rightarrow
\infty} \{ 1/2^{{\kern 1pt}i}\}
  $$
is the initiator, $\varrho$ the self-similarity ratio and a the
additional number of subfigures inserted in the $(1/\varrho)$
fragments of the initial figure. The corresponding fractal
structure is denoted ($\Gamma$). It has been shown in Part 2 that
($\Gamma$) can be decomposed in a sequence of elementary
components \{C1, C2,..., Ck,...\}. If all these elementary
deformations are gathered on one single ball, then this ball
contains all the quantum of fractality, though its dimension is
not changed. It is therefore nonmassive as it stands and its
motion is determined by the velocity of transfer of nonmassive
deformations, that is the maximum permitted by the elasticity of
the space lattice. Since the deformations are ordered and
distributed in one particular structure, it owns a stability
through mappings of Poincar\'{e} sections. Such particles are
likely to correspond to bosons, that is to pseudoparticles
representing just transfer of packs of deformations in a isolated
form.

Hence, photon-like corpuscles will carry the equivalent of various
quanta of fractality $\rm \{\vartheta_{{\kern 1pt}i}\}_i$, that
is, their equivalent in mass in a decomposed form. This represents
as many deformations of the lattice, and finally of equivalent in
energy.

\medskip \noindent
\textbf{3.3.1.2. Families of massive particles.} Any single ball
carrying a group $\rm \{ \Gamma_{{\kern 1pt}i} \}_i$ of quantum
fractals will represent a class of massive particles. Depending on
both the number and the mode of association of these fractal
quanta, various symmetries will result and provide these classes
with specific properties.

Hadron-like families will thus be represented by the following common
structures:
 $$
\rm H_{{\kern 1pt}i,k} = \big(\{a_{{\kern 1pt}i}, {\kern
1pt}\varrho_{{\kern 1pt}i}, {\kern 1pt}e_{{\kern
1pt}i}\}_{i,k}\big)
 \eqno(5.1)
 $$

Simpler particles made from one single quantum of fractality $\rm
\{\varrho, {\kern 1pt} a\}$ would likely correspond with
lepton-like structures, such that:
 $$
\rm L_{{\kern 1pt}i} = \big( \{ N_{{\kern 1pt}i} \cdot
(\varrho_{{\kern 1pt}i}){\kern 1pt}^{e_i}\}\big).
 \eqno(5.2)
 $$

\subsubsection*{3.3.2. Spins for hadron-like balls}

\textbf{3.3.2.1. Fermion-like cases.} Moving massive balls have
been shown to carry a cloud of deformations transferred to
degenerate balls of the surrounding space, with periodic exchange
between this "inertons" cloud and the original particle (Part 2).
The period of this pulse has been identified with the de Broglie
wavelength. Hence, the center of mass (y) of the system composed
of the particle and the inerton cloud permanently undergoes a
movement forwards and backwards along the trajectory of the
system. Two canonical positions are possible, with respect to the
particle: \\

\noindent (i) y is centered on the particled ball, and

\noindent (ii) t is no longer centered. \\

The probability of state of x is thus P(y) = 1/2.

\medskip \noindent
\textbf{3.3.2.2. Boson-like cases.} Consider a ball carrying
quanta of masses in the decomposed form: then, such a system is
opposed the minimum resistance by the surrounding degenerate
balls, which are of the same nature, excepted that their
individual densities of deformation are much smaller. Therefore,
boson-like particles do not generate a cloud similar to that of a
massive particle, and their center of mass (y) owns only one main
state: thus P(y) = 1.

\medskip \noindent
\textbf{3.3.2.3. Spin module.} The state of the center of mass is
assessed by the expected moment of junction $\rm \langle MJ
\rangle$ of its components, so that the spin-like system is
described by $\rm P(x,y) \cdot\langle M \rangle$, standing for
$\rm s \cdot \hbar/2$, that is, the classical spin module
expression.

This parameter would likely be summable over an association of
particles into a more complex system, which is consistent with the
additivity of spins.

In all cases eddy-like components of the motion concern the
relative behavior of the particle and of its inertons cloud,
respectively. These relative rotation movements could likely be of
opposite sense, and at least in some cases under current
investigation, the whole \{particle + inertons\} system may either
escape rotation, or get a resulting rotation axis and speed,
depending on the rotation parameters of the most massive part of
the system. Then, a rotation pulse with reversion of direction can
be expected in some conditions.

\subsubsection*{3.3.3. Charges}

\textbf{3.3.3.1. Opposite kinds of particle deformations.} When a
quantum of fractal deformations collapses into one single ball,
two adjacent balls exhibit opposite forms: one in the sense of
convexity and the other in the sense of concavity. Hence, there
occurs a pair instead of a single object. The paired structures
hold the same fractal dimension, and they will retain the same
masses if they get the same volumes. This is realized if the
member of the pair whose deformation is in the convex sense looses
an equivalent volume in a nonfractal form, as schematically shown
in Figure 1.

\begin{figure}
\begin{center}
\includegraphics[scale=1.6]{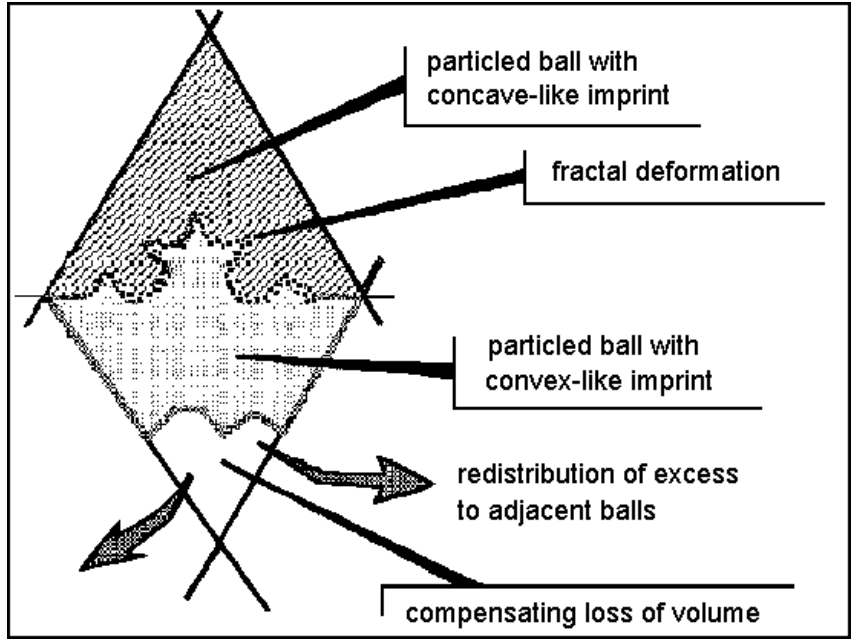}
\caption{\small{Schematic representation of the paired
deformations produced by the collapse of a quantum of fractal
deformations into a ball. One of the two complementary topologies
could be called a positive charge and the other a negative one.}}
\label{Figure 1}
\end{center}
\end{figure}

The progression of such structures in the degenerate space will
generate several kinds of inerton cloud equivalent, depending on
convexity trends ($\Xi$) and symmetry properties ($\Psi$) of the
corresponding structures.

The properties generated by $\rm Q = \{\Xi, {\kern 1pt}\Psi \}$
will be called "charge effects".

\medskip \noindent
\textbf{3.3.3.2. Electric and magnetic charges.} The motion of a
particled ball in the space lattice exhibits some similarity with
a cutwave which would be produced by a boat made with water. The
shape of the waves depend on the shape of the moving object, as
shown in Figure 2.

\begin{figure}
\begin{center}
\includegraphics[scale=1.6]{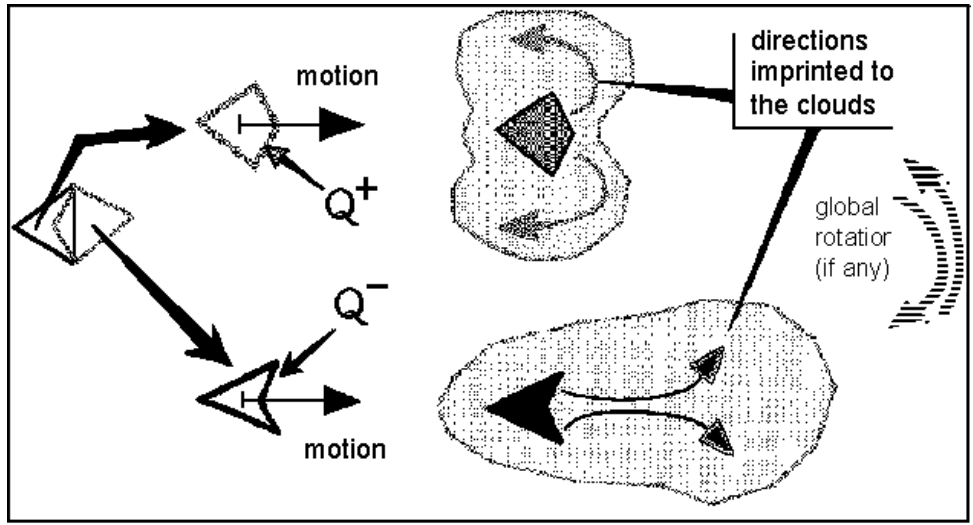}
\caption{\small{Changes imprinted in the shape of the inerton
clouds by the shape of the particles deformations of symmetric
type.}} \label{Figure 2}
\end{center}
\end{figure}

\begin{figure}
\begin{center}
\includegraphics[scale=1.6]{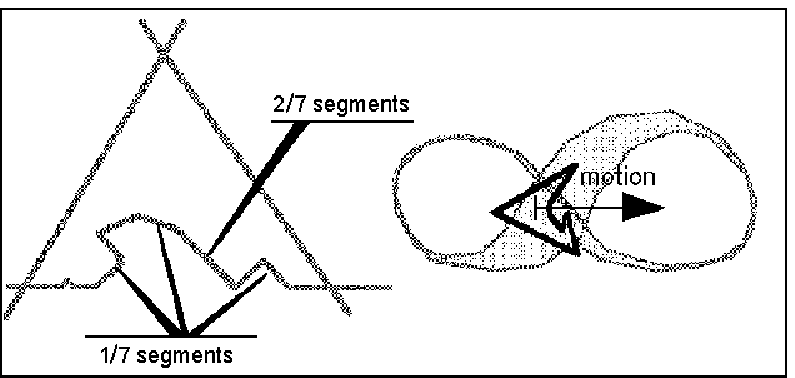}
\caption{\small{An asymmetric component in the deformation of a
particle induces a torsion of the inerton cloud. \newline
\noindent {\bf Note:} The figure illustrates a initiator figure
composed of seven subsegments of ratio (1/7) and two subsegments
of ratio (2/7) of the modified side. Thus, the fractal is roughly
depicted by $\rm 7 \cdot (l/7){\kern 2pt}e + 2\cdot (2/7){\kern
2pt}e = 1$.}} \label{Figure 3}
\end{center}
\end{figure}

Component $\{ \Xi \}$ of Q produces two kinds of inerton clouds
(Figure 2), which can be identified with electric fields, since
they depart from the "neutral" inerton cloud by a symmetric kind
of deformation. These shapes are complementary and can be
associated in a oriented field, providing the corresponding area
of the lattice with vectorial properties.

Component $\{ \Psi \}$ appears when the fractal clusters are no
longer symmetric in a massive particle. This case corresponds to
well characterizable parametrizations. For instance, a sufficient
condition is that the fractal quanta of masses have the following
form, where at least some $\rm N_{{\kern 1pt}i}$ are odd numbers:
 $$
\rm (\Gamma) = \Big( \{ N_{{\kern 1pt}i}, {\kern 1pt} a_{{\kern
1pt}i}, {\kern 1pt} \varrho_{{\kern 1pt}i} \}: {\kern 2pt}{\kern
2pt} {\mathop \sum_{i} }{\kern 2pt}{\underline{N}}_{{\kern
2pt}i}\cdot (\varrho_{{\kern 1pt}i})^{{\kern 0.5pt} e_{{\kern
0.5pt}i}}=1 \Big)
 \eqno(5.3)
 $$

In such cases, shown by Figure 3, the asymmetry will provide the
cloud of inertons with an additional deformation represented by a
torsion.

 Neutrality may not be so simple. In a strict sense,
it can be reflected by the absence of deformation. However, it can
also be represented by symmetry and homogeneity of the
distribution of convex and concave components simultaneously
present on the same edges of particles, while the volume reduction
associated to mass would remain fulfilled. The latter case would
stand for a pseudo-neutrality worth to be taken in consideration.
Figure 4 illustrates these features in a quasi-metaphoric sense.

\begin{figure}
\begin{center}
\includegraphics[scale=1.6]{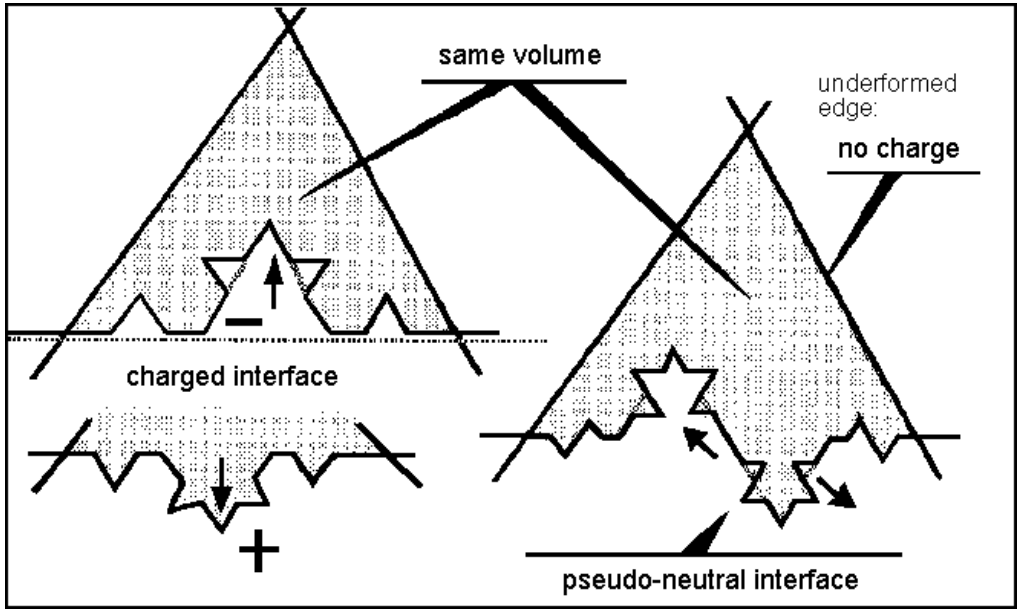}
\caption{\small{Illustration of canonical particles differing only
by their charge. A strictly neutral edge would have no
deformation. A charged edge can show cavities or protuberances.
\newline  \noindent  {\bf Note:} A pseudo-neutral interface
would have the same quantity of cavities and protuberances and it
would be in theory dissociable into two opposite charges.}}
\label{Figure 4}
\end{center}
\end{figure}

\subsubsection*{3.3.4. Predicted expansion and the "quintessence"}

\textbf{3.3.4.1. Introduction.} Each time the distribution of
degenerate deformations collapses into particled balls, there
occurs a corresponding increase of volume of the surrounding
balls, which compensates the reduction of volume in the particled
cell. The motion of the particle can likely be provided by the
reaction to the creation of this kind of "anti-inerton cloud,
which behaves in a opposite way than the resistance of the inerton
cloud to the motion of the particle.

These increases of volume are then progressively scattered by transfer of
the corresponding deformations to an expanding cloud of "dilatation" quanta.
This phenomenon operates a gain of space volume away of the particles:
therefore, it represents a kind of force acting in a way opposite to the
gravitation. This suggests two main corollaries.

\medskip \noindent \textbf{3.3.4.2. Quintessence.}
The existence of a "fifth cosmic element", somewhat related with
Einstein's "cosmological constant" has been thoroughly discussed
(see Krauss, 1999; B\"orner, 2000; Ostriker and Steinhardt, 2001,
etc. for review). In our model, this factor appears strictly in
connection with the creation of matter from the degenerate
lattice, which may stand for the cosmic form of a void. Its
quantitative expression is directly correlated with the density of
fractal deformations, that is of energy, and its range will be
shown below to be of the long type. Basically, the above theorem
of distribution shows that it appears independently of any
previous presence of matter nor radiations. Last, it seems not to
be braked by gravitational forces, since it appears as a
by-product of the same event which produces gravity. All these
points make this compensatory dilatation phenomenon a candidate
for the "quintessence".

\medskip \noindent \textbf{3.3.4.3. Space expansion.}
The transfer of elementary volumes released by the formation of
massic particles occurs within the frame of the degenerate space,
since it is processed in a nonfractal state, just for the
compensation of lost volumes, without need for dimensional change.
Call (a) such a element of volume: first, it is finite.

While the area of the iterated self-similar transform is
theoretically infinite, its volume in 3-D sections is not.
Therefore, while a part of the volume of the transformed cell is
reduced by a finite value, the volume of surrounding cells is
increased by a corresponding finite value. The homogeneity of the
lattice is partly restored through progressive transfer of the
additional volume to neighbor cells.

\medskip \noindent
\textbf{Remark 3.4.} A degenerate ball constitutes a
"superparticle".

\medskip \noindent \underline{Proof.} The oscillations of an elementary
cell have been considered by Krasnoholovets (1997; 2000) as the
"degenerate" state of space. It is a state without formation of a
particle, though potentially able to provide particles upon
proposition 3 and statement 3. One elementary ball thus
constitutes the putative generator of particles, what has been
called a "superparticle" in previous attempts for unification of
theories.

\medskip \noindent
\textbf{Remark 3.5.} In any one Poincar\'{e} section, representing
a timeless instantaneous state (an instans) of universe, the
lattice of space is represented by a stacking of balls with
nonidentical shape.

\medskip \noindent
\textbf{Conjecture 3.1.} Elementary balls exhibit increasing
volumes from the center to the periphery of a 3-D stacking.

\medskip \noindent \underline{Justifications.}  Three arguments
concur to the same proposition. \newline (i) Oscillating
deformations in excess in one cell can be partly compensated by
transfer to neighboring balls, like an equivalent to the inerton
cloud surrounding a particled ball. However, in central parts, the
volume available is limited by the density of the stacking, and
this limit is likely decreasing while going to the outer coats of
the lattice. In a simple estimation, we denote by (a) the radius
of the canonical (smallest) volume which can be transferred from a
ball to another. Assuming that each cell forwards a volume (a) to
another situated closer to the periphery, in the stacking, then
the radius of a ball in the n$\rm ^{th}$ coat is approximated by:
 $$
\rm r_{{\kern 1pt}n} = r_{{\kern 1pt} i} + (n-1){\kern 1pt}a
 \eqno(5.4)
 $$
\newline
(ii) While the above considerations are valid for a particleless
lattice, if the lattice is filled with particled balls, then there
results a kind of pressure due to the inerton clouds. Hence,
relation (4.2) is affected a corrective quantitative term to (a)
and its distribution is determined by the distribution of
particled balls in the considered space.
\newline
(iii) In contrast with the finiteness of volume to be
compensatively distributed in the surrounding cells, the area of a
particled cell is infinite, and the needed area cannot be
compensated by a finite number of the surrounding cells. Thus an
influence of any particle is likely to be found up to the most
remote parts of the lattice.

The last two points will be further examined more in details in the third
part of this study, through involvement of the concept of quantum of
fractality in relation with mass of particled cells.

\medskip \noindent
\textbf{Corollary 3.2.} Since elementary balls can be found at
various scales, due to the quantic ratios which characterize the
lattice, as shown above, this means that elementary particles are
not of one unique size.

\medskip  \noindent
\textbf{Corollary 3.3.} Transfers of non fractal elementary
volumes between balls are operated without dimensional increase.

\medskip \noindent
\underline{Proof.} At each given scale, the corresponding
increments (a) are represented by similar topological features. In
effect: following relation (3), we have for $\rm n=2$: $\rm r_2 =
r_1 + a$, and for $\rm n=1$: $\rm r_1 = r_0 + (n-1){\kern 1pt}a =
r_{0}$. Since then $\rm r_0 = r_2 - a$, $\rm r_0$ stands for a
founding ball. Let $\rm r_0 =${\kern 3pt}{\O}, an empty set. Then,
$\rm r_2 = r_0 + a$ can be represented by ({\O}, \{{\O}\}) where
\{\O\} is the frontier of ball r$_2$. The element \{\O\} is what
is exchangeable, and since it is a frontier, it has a dimension
lower than the dimension of the interior, that is: $\rm dim (a) <
dim (r)$. Thus exchanges do not modify the dimensionality of
involved balls.

However, mass transfers from a particled cell to its surrounding
degenerate balls involves a distribution of quanta of fractality
through the concept of fractal decomposition described above.

\medskip \noindent
\textbf{Remark 3.6.} Consequently, it may be considered that these
exchanges apply to the frontiers of the balls, which will result
in changes in the density of their internal structures. It is
noteworthy that the density has been used as a probe for the
identification of the packing of balls, though in this case only
solid balls are considered (Hales, 2000). Otherwise, the
adjunction of (a) to (r) may result in the reunion of two spaces
having nonequal dimensions, which can result in a structure of the
"beaver space" type as described in Part 1.

\medskip \noindent
\textbf{Corollary 3.4.} A measure on such a lattice space by using
a scanning function as described in Part 1 will not scan the same
components in elementary balls situated at various distances from
its origin. Since there likely occurs an increase of the
composition of balls from this origin, then the gauge will
decrease with increasing distances: in effect, a larger set of
scanned structures will appear at farther distances. Then, remote
distances will be overestimated by a measure using a local gauge.
This might account for the phenomenon known as the Doppler effect,
in turn usually involving the Hubble constant. It should be noted
here that the interpretation of the redshift has been matter of
diverging treatments (e.g.: Hannon, 1998).

\subsubsection*{3.3.5. Towards a formalism to Big-Bang(s)}

\hspace*{\parindent} In Part 1 of this study, it has been proved:

\noindent (i) that the lattice existing from empty hyperset units
provide a manifold of quantic scales represented by a set of
defined integer ratios (Bounias and Krasnoholovets, 2002a) and

\noindent (ii) that there exists empty set units of various size,
with integer vs.rational similarity ratios.

One universe $\rm U_{ j} = \{\{S_{{\kern 1pt}i}\}_i, {\perp}
\}_j$, represented by one particular sequence of Poincar\'{e}
sections selected through a particular combination rule is nothing
but a manifold of organized empty set units, and since the lattice
in which it is embedded is strictly fractal, the reunion of these
empty set units is a higher scale empty set. Thus, $\rm U_{j} =
${\O}$_j$.

Now, consider the part of the embedding lattice in which
{\O}$_{{\kern 1pt} j }$ owns just the size of a free ball. This
"over universe" denoted {\O}$_{+ j }$ will behave like described
through Proposition 1. The distribution collapse of degenerate
balls of {\O}$_{+ j}$ will result in the formation of a particle
whose subparts contain potentially as many quanta of fractality as
$\rm U_{j}$ contains massive objects. Thus, what represents a
creation of a particle inside $\rm U_{j}$ is a primordial
condensation of a ball into $\rm U_{j}$ inside $\rm U_{+j}$.

This suggests that such kinds of "Big-Bangs" may have occurred, occur and
will occur in at least denumerably many balls of the embedding lattice,
without need for a "outside" provision of energy.

However, these "Big-Bangs" fulfill some conditions. In effect, it
has been specified in Part 1 of this study that universe is
definitely constructed in a specified space ($\rm W^{\omega}) =
\{X, \perp \}$ with set $\rm X \in \{X^{3} \cap X^{4}$) and
combination rule ($\perp$) determining the choice of S{\kern
0.3pt}i{\small{+1}} from Si. Hence, relations between different
universes and past-to-present successive universes can exist only
through the same law ($\perp$).

\section*{4. Discussion and Conclusions}

\hspace*{\parindent} The law ($\perp$) proposed as the operator of
the selection of successive Poincar\'{e} sections constituting a
spacetime presents the interest, besides providing continuity of
this spacetime, of keeping inert or low-moving structures (like
mountains, landscapes, etc.) stable. Furthermore, it brings as a
corollary that events will basically follow the shorter path
between two steps, which is consistent with both the least action
principle and the geodesic trajectory principle. This suggests
that the kind of universe that we have described is consistent
with our observable spacetime, even if our description of
submicroscopic events, through the formalism of set theory
extended to nonwellfounded sets (a consistent extension) may be
considered in some sort as a metaphoric description.

The components of the \{particle + inertons\} system are likely
inhomogeneous as topological balls do not need to be strictly
spherical (the latter case is just a particular one).

Therefore, their coexistence in a single system representing the
dual \{wave/particle\} system deserves special attention, since
spin-related properties could reflect the eddy-like motion that
inhomogeneity should impulse to the components and finally, in a
resulting manner, to the system. Such properties have been
described by Lin and OuYang (1980; 1998), Wu and Lin (2002), while
Lin (1988) explored the compatibility of world exploration with
the theoretical study of systems: these goals are well converging
with our objective of mathematical exploration of an unknown
world, as developed in Part 1 of this study.

The theoretical reasoning presented in this study, following the
basis developed in Parts 1 and 2, sheds some light on the question
of the hypothetic "origins" of universe. In fact, there is no need
for beginning nor for end. Even the expansion might not induce the
consequences expressed through other approaches in terms of
forever expansion and progressive immobilization, nor cyclic
contraction and collapse. Our approach, basically founded on a
formal justification of existence of "something", and then on
corpuscular description and properties, turns to introduce some
insights about cosmic scales and cosmic-size properties.
Interestingly, Andre\"{\i} Linde pioneeringly suggested that a
"Grand-Universe" could be composed of bubbles of universes that
could form and disappear in various parts in an independent
fashion. Though it was not primarily our aim to treat these
questions, it turns out that the development of our model from
defined startpoints comes to support Linde's hypothesis.

Furthermore, the former hypothesis raised long ago by Feynman
about particle trajectories which would be infinite and not
derivable is consistent with our proposition that particles are
distinguished from the degenerate space by a shift of dimensional
properties, that is with a fractal organization.

The next part of this study will aim to examine more in detail
what are the peculiarities of the various kinds of corpuscles
predicted by our approach.

\bigskip

\section*{References}

Bonaly, A., Bounias, M., 1995. "The trace of time in Poincar\'{e}
sections of topological spaces", {\it Physics Essays}, Vol. 8, No.
2, pp. 236-44.  \\

\noindent B\"{o}rner, G., 2000. "The infinitely large", {\it Pour
La Science} ({\it Scientific American}, French edition), Vol. 278,
pp. 120-7.   \\

\noindent Bounias, M., Bonaly, A., 1997. "Some theorems on the
empty set as necessary and sufficient for the primary topological
axioms of physical existence", {\it Physics Essays}, Vol. 10, No.
4, pp. 633-43.   \\

\noindent Bounias, M., Krasnoholovets, V., 2003a. "Scanning the
structure of ill-known spaces: Part 1. Founding principles about
mathematical constitution of space", {\it Kybernetes: The
International Journal of Systems and Cybernetics}, Vol. 32, No.
7/8, pp. 945-75. (Also physics/0211096.)   \\

\noindent Bounias, M., Krasnoholovets, V., 2003b. "Scanning the
structure of ill-known spaces: Part 2. Principles of construction
of physical space", {\it Kybernetes: The International Journal of
Systems and Cybernetics}, Vol. 32, No. 7/8, pp. 976-1004.
(Also physics/0212004.)   \\

\noindent Frizsh, H., 2000. "The infinitely small in physics",
{\it Pour La Science} ({\it Scientific American}, French edition),
Vol. 278, pp. 112-9.   \\

\noindent Hannon, R.J., 1998. "An alternative explanation of the
cosmological redshift", {\it Physics Essays}, Vol. 11, No. 4,
pp. 576-8.  \\

\noindent Hales, T.C., 2000. "Cannonballs and honeycombs", {\it
Notices of the AMS}, Vol. 47, No. 4, pp. 440-9.  \\

\noindent Krasnoholovets, V., 1997. "Motion of a relativistic
particle and the vacuum", {\it Physics Essays}, Vol. 10, No. 3,
pp. 407-16. (Also quant-ph/9903077). \\

\noindent Krasnoholovets, V., 2001. "On the way to submicroscopic
description of nature", {\it Indian Journal of Theoretical
Physics}, Vol. 49, No. 2, pp. 81-95. (Also
quant-ph/9908042).    \\

\noindent Krauss, L., 1999. "Antigravity", {\it Pour La Science}
({\it Scientific American}, French edition), Vol. 257, pp. 42-9.   \\

\noindent Lin, Y., OuYang, S.C., 1996. "Exploration of the mystery
of nonlinearity", {\it Research of Natural Dialectics}, Vol.
12-13, pp. 34-7.  \\

\noindent Lin, Y., 1988. "Can the world be studied in the
viewpoint of systems?", {\it  Math. Comput. Modeling},
Vol. 11, pp. 738-42.  \\

\noindent  Lin, Y., OuYang, S.C., 1998. "Invisible Tao and
realistic nonlinearity propositions". {\it Kybernetes: The
International Journal of Systems and Cybernetics}, Vol. 27, pp. 809-22.  \\

\noindent Ostriker, J., Steinhardt, P., 2001. "The fifth cosmic
element". {\it Pour La Science} ({\it Scientific American}, French
edition), Vol. 281, pp. 44-53.  \\

\noindent R\"{u}egg A., 1985. {\it Probabilities and Statistics},
Presses Polytechniques Romandes, Mausanne, Switzerland, pp. 52-87.   \\

\noindent  Wang, L., Caldwell, R., Ostriker, J., Steinhardt, P.,
2000. "Cosmic concordance and quintessence", {\it Astrophysical
Journal}, Vol. 530, No. 1, pp. 17-35.   \\

\noindent  Wu, Y., Lin, Y., (2002). "Beyond nonstructural
quantitative analysis", in {\it Blown ups, spinning currents and
modern science}, World Scientific, New Jersey, London, p. 324.  \\

\bigskip

\subsection*{Further reading}

Caldwell, R., Dave, R., Steinhardt, P., 1998. "Cosmological
imprint of an energy component with general equation of state",
{\it Phys. Rev. Lett.}, Vol. 80, No. 8, pp. 1582-5. \\

\noindent  Krauss, L., 1998. "The end of age problem, and the case
for a cosmological constant revisited", {\it Astrophysical
Journal}, Vol. 501, No. 2, 461-6. \\

\noindent Schwartz, L., 1997. {\it Un math\'{e}maticien aux prises
avec le si\`{e}cle}, Odile Jacob, Paris, p. 250.

\end{document}